# A Generalization of the Concept of Parity Conservation

J.N. Pecina-Cruz

IAT The University of Texas at Austin
Austin, Texas 78759-5316
jpecina@intelligent-e-systems.com

This article rebuilds the parity conservation under the light of Heinsenberg's uncertainty principle and the equivalence among the four space-time coordinates. The lack of equilibrium between matter and antimatter is elucidated; neutrino mass is a logical corollary of this asymmetry. Based on this new concept of parity conservation the existence of black holes is questioned. It is also found that entropy could decrease in a close quantum system. CPT would be an exact symmetry if the "full" Poincare group were the symmetry group of nature. Therefore, CPT must be violated if nature obeys a larger symmetry group than that of the "full" Poincare group. Examples of CPT violation could be found in a quantum theory of gravity, and the supertheories.

**Introduction**

Symmetries in nature have played a significant role in the development of new ideas in physics. Maxwell displacement current was derived as an attempt to maintain the symmetry between the magnetic and electric fields in his famous Maxwell's equations [1]. Similarly, Einstein's theory of relativity demands to be a symmetry that leaves invariant the laws of physics [2] under the relative motion of reference frames. The eightfold way symmetry of Ne'eman and Gell-Mann [3] constitutes another achievement of the human thought. It would be striking for an exterior intelligence to observe the left and right symmetry almost everywhere in the universe. This was the reason why the scientific community was surprised with the radical interpretation of the theta-tau puzzle from C.N. Yang and T.D. Lee [4]. L. Landau pointed out that a solution to save the parity symmetry is generalizing it, by incorporating the interchange of particles by antiparticles [5] in a spatial mirror reflection. L. Landau claimed CP was the correct symmetry for nature. Similarly, C.N. Yang recognizes that if the parity symmetry is generalized by incorporating the exchange of particles by antiparticles, then parity is conserved [6]. However, it has been found that CP is not, in general, a symmetry transformation [7], but CPT [26] is. The authors of reference 7 complain about the lack of a more profound understanding of CP violation, and they believe that parity symmetry must be restored. C.N. Yang accepts the possibility that he might be misinterpreting the theta-tau puzzle by claiming left and right symmetry violation [6]. To illustrate his concerns he quotes an apparent violation in the symmetry of the magnetic field in an example given by E. Mach. But, C.N. Yang could not find a rational justification to the substitution of matter by antimatter in a spatial mirror reflection [6]. A fragment of Yang's meditations is included in this article:

"…the question remains why it is necessary in order to have symmetry, to *combine* the operation of switching matter and antimatter with a mirror reflection. The answer to such a question can only be obtained through a deeper understanding of the relationship between matter and antimatter. No such understanding is in sight today."

In this paper a logic scheme is proposed that explains why a particle must be replaced by its antiparticle in a spatial mirror reflection. In the following sections, it will be proven that a spatial mirror reflection followed by the exchange of a particle by its antiparticle is the result of reflecting the four space-time coordinates.

**1. Negative Energy States**

Misleading concepts such as the existence of a ether cosmic or the belief that heavy objects fall faster than light ones, plagues the history of physics. At the same time nature shows us secrets that sometimes we are reluctant to accept. This might be the case of E. Schrödinger when he renounces to a relativistic formulation of the wave equation because the appearance of negative energy quantum states [8]. It was P.M.A. Dirac that stepped further introducing his hole-theory [9]. Dirac's hole theory not only interpreted quantum negative energy states, but also gave a reality to the vacuum.

If the Landau's generalization of conservation of the parity is correct, why does matter have to be substituted by antimatter in a mirror reflection? This question can be answered under the light of Einstein's time interpretation. In contradiction with our common sense, Albert Einstein astounded the scientific world by interpreting time as another coordinate in a continuous of four-dimensions and on equal footing than the space coordinates. *Parity is violated because time is not considered as an additional coordinate in the inversion procedure. Therefore, an incomplete symmetry is accomplished.* Time reversal symmetry does not have any different meaning than that of the coordinate x (y or z) reversal. Time does not have any special privileges over their counterparts; the space coordinates. Time can be contracted or enlarged just like any space coordinate. The theory of relativity is based on the assumption of the laws of physics are independent of the coordinates' reference frame, in which time coordinated is embedded. This significant lack of time coordinate privileges is not incorporated in C.N. Yang's and T.D. Lee's concept of the reflection symmetry. Therefore, special attributes are given to this coordinate over its counterparts, the space coordinates; time again is placed on a special stadium, as it was in the physics pre-relativistic. Therefore, an incomplete symmetry is achieved, which is violated in some physical processes, such as in the weak interaction. Similar symmetry violation would take place if we imposed reflection invariance for the x coordinate alone in Maxwell's equations.

It seems that the reason to abandon a simultaneous reflection of the four coordinates (strong reflection [10]) is the introduction of the quantum negative energy state of a particle. This negative energy state is obtained when time is reflected [11]. In spite of this rejection to incorporate negative energy states in physics, the need in nature of those states motivated Dirac [9], Schwinger [12], Pauli, Weisskopf [13], Feynman, and Stückelberg [14] to formulate a field theory. Dirac, Feynman and Stückelberg formulated a single particle field theory, while Pauli, Schwinger and Weisskopf formulated the many particles field theory. Under Feynman-Stückelberg's single particle field theory, parity symmetry is recovered, as it will be discussed below. The strongest argument against the negative energy states rests on the assumption that the elementary particles that obey the Bose-Einstein statistic would fall down to the lowest negative energy state. *Fermi-Dirac and Bose-Einstein statistics are not valid for distances shorter than the Compton's wavelength of an assembly of particles.*. Since Heinsenberg's uncertainty principle [15] limits the range of validation of these statistics, whether or not two identical particles of Bose, separated by their Compton wavelength, can occupy the same quantum state is uncertain, because the validation of the statistics is conditioned by the Compton wavelength of the particles. Then, such collapse is not justified in the boson case. Additionally, the Pauli's exclusion principle argument is untenable in the Dirac's hole-theory for distances less than the Compton wavelength of the particle. Therefore, a new interpretation must exist; this interpretation has to place both *boson and fermions under the same theoretical frame*. In the next sections such scheme is proposed by introducing novel arguments.

**2. Matter-Antimatter and the Violation of Causality**

In the one particle scheme Feynman and Stückelberg interpret antiparticles as particles moving backwards in time [14]. This argument is reinforced by S. Weinberg who realizes that antiparticles existence is a consequence of the violation of the principle of causality in quantum mechanics [16]. The temporal order of the events is distorted when a particle wanders in the neighborhood of the light cone. How is the antimatter generated from matter? According to Heisenberg's uncertainty principle a particle wandering in the neighborhood of the light-cone suddenly tunnels from the timelike region to the spacelike region; in this region the relation of cause and effect collapses. Since if an event, at $x_2$ is observed by an observer A, to occur later than one at $x_1$, in other words $x_2^0 > x_1^0$. An observer B moving with a velocity **v** respect to observer A, will see the events separated by a time interval given by

$$x'^0_2 - x'^0_1 = L^0_\alpha(v)(x_2^\alpha - x_1^\alpha), \qquad (1)$$

where $L^\beta_\alpha(v)$ is a Lorentz boost. From equation (1), it is found that if the order of the events is exchanged for the observer B, that is, $x'^0_2 < x'^0_1$ (the event at $x_1$ is observed later than the event at $x_2$.), then a particle that is emitted at $x_1$ and absorbed at $x_2$ as observed by A, it is observed by B as if it were absorbed at $x_2$ before the particle were emitted at $x_1$. The temporal order of the particle is inverted. This event is completely feasible in the neighborhood of the light-cone, since the uncertainty principle allows a particle tunnel from time-like to space-like cone regions. That is the uncertainty principle will consent to the space-like region reach values above than zero as is shown in next equation,

$$(x_1 - x_2)^2 - (x_1^0 - x_2^0)^2 \leq \left(\frac{\hbar}{mc}\right)^2,$$

where $\left(\frac{\hbar}{mc}\right)^2 > 0,$ (2)

and $\frac{\hbar}{mc}$ is the Compton wave-length of the particle. The left hand side of the equation (2) can be positive or space-like for distances less or equal than the square of the Compton wavelength of the particle. Therefore, causality is violated. The only way of interpreting this phenomenon is assuming that the particle absorbed at $x_2$, before it is emitted at $x_1$ as it is observed by B, is actually a particle with negative mass and energy, and certain charge, and spin, moving backward in time; that is $t_2 < t_1$ [16]. This event is equivalent to see an antiparticle moving forward in time with positive mass and energy, and opposite charge and spin that it is emitted at $x_1$ and it is absorbed at $x_2$. With this reinterpretation the causality is recovered.

The square of the mass for the observers A and B is a Poincare invariant. In other words, *the mass of a particle itself is not an invariant, but its square* [28]. There is nothing to prevent that the rest mass could have different sign in distinct reference frame, as happen with its energy, electric charge and spin. In this paper, it is conjectured that a particle moving backward in time possesses a negative mass [16]. When it is observed as an antiparticle moving forward in time its mass is positive. Still its mass square is a Poincare invariant.

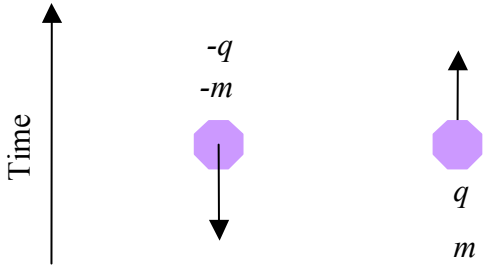

Figure 1. A particle moving backwards in time is equivalent to an antiparticle moving forward in time with opposite spin (not showed), charge, and mass.

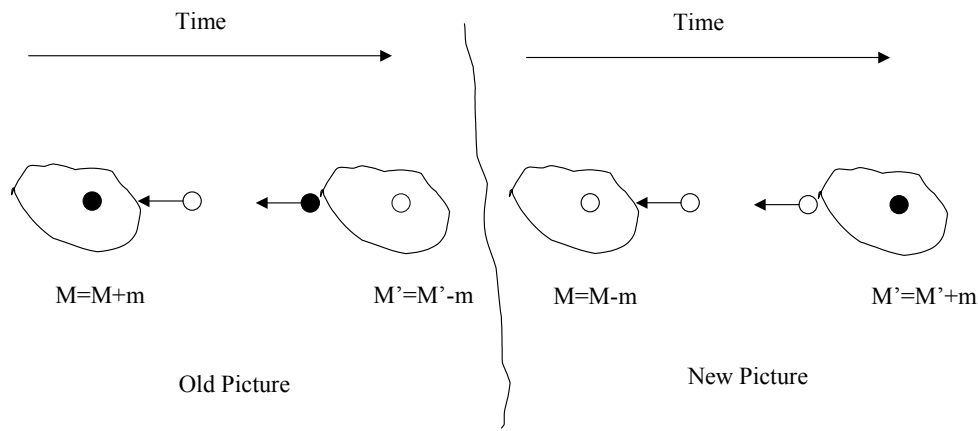

Figure 2. Exchanging a particle moving backwards in time

In the left part of Figure 2, the current interpretation is shown for the emission and absorption of an antiparticle. A particle of positive mass moves backwards in time. The particle left a hole of mass m, in M' which it is later absorbed by M. In the right part of Figure 2, it is illustrated the new interpretation, given in this paper, a particle of negative mass moves backward in time leaving a hole of negative mass -m, in M', which is equivalent to insert a positive mass +m, in M'. The mass M absorbs a negative mass –m, then M decreases by m units its mass.

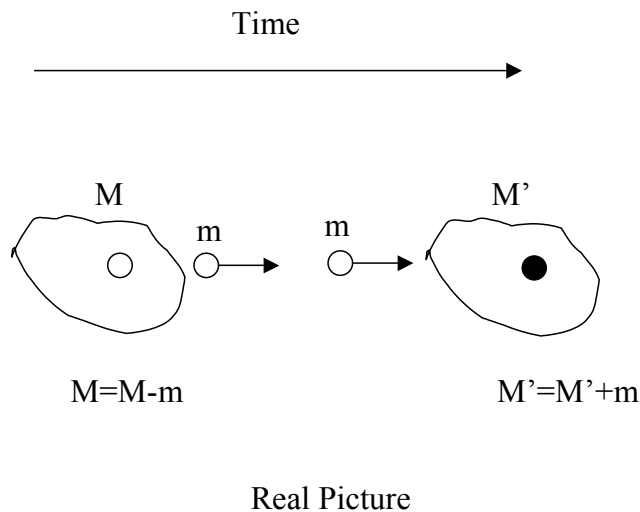

Figure 3. Exchange of an antiparticle of positive mass moving forward in time

In Figure 3, it is shown the equivalent phenomenon to the one depicted on the right part of Figure 2. An antiparticle of positive mass, moving forward in time, is emitted by mass M and absorbed by mass M'.

An argument due to Dirac [17], and Fock [18]; which has been augmented by the author of this article illustrates this conjecture. Let us consider a charge particle under the influence of an exterior electromagnetic field. And let us study the negative energy solutions of the particle wave equation. In the matrix Dirac's representation if we exchange the Dirac's matrices by $\alpha_1^0 = \alpha_1, \alpha_2^0 = \alpha_m, \alpha_3^0 = \alpha_3, \alpha_4^0 = \alpha_2$, a new representation is obtained with matrices $\alpha_1^0, \alpha_2^0, \alpha_3^0$ real, and $\alpha_4^0$ pure imaginary or zero matrix.

$$\left(i\hbar\frac{\partial\psi}{\partial t} + \frac{e}{c}A_0\psi\right) - \alpha_1^0\left(i\hbar\frac{\partial\psi}{\partial x} + \frac{e}{c}A_x\psi\right) - \alpha_2^0\left(i\hbar\frac{\partial\psi}{\partial y} + \frac{e}{c}A_y\psi\right) -$$

$$\alpha_3^0\left(i\hbar\frac{\partial\psi}{\partial z} + \frac{e}{c}A_z\psi\right) - mc\alpha_4^0\psi = 0 \quad (3)$$

If this equation is conjugated and its coordinates reflected (x,y,z,t) →(-x,-y,-z,-t), and assuming Maxwell's equations invariance, that is, that the fields transform according to [19]

$$\vec{E} = -\frac{1}{c}\frac{\partial\vec{A}}{\partial t} - \vec{\nabla}\Phi \quad (4)$$

Where **E** is the electrical field, **A** is the vector potential and $\Phi$ is the scalar potential. From the invariance of equation (4) under space-time reflections (x,y,z,t)→(-x,-y,-z,-t), we get that **A** →-**A**, $\Phi$ →-$\Phi$. After substituting in equation (3), equation (5) is obtained.

$$\left(i\hbar\frac{\partial\overline{\psi}}{\partial t} - \frac{e}{c}\Phi\overline{\psi}\right) - \left(i\hbar\frac{\partial\overline{\psi}}{\partial x} - \frac{e}{c}A_x\overline{\psi}\right)\alpha_1^0 - \left(i\hbar\frac{\partial\overline{\psi}}{\partial y} - \frac{e}{c}A_y\overline{\psi}\right)\alpha_2^0 \quad (5)$$

$$- \left(i\hbar\frac{\partial\overline{\psi}}{\partial z} - \frac{e}{c}A_z\overline{\psi}\right)\alpha_3^0 + mc\overline{\psi}\alpha_4^0 = 0$$

Therefore, if in equation (3) are replaced e by –e, m by –m, and the wave function by its conjugate complex, then the equation of the charged particle in a negative energy quantum state moving backward in space-time is obtained. The replacement of the function $|\psi\rangle$ by its conjugate complex $|\overline{\psi}\rangle$ is induced by the change of direction of time. This point will be discussed in the next section, when the unitary irreducible representations of the Poincare group with simultaneous space-time reflections will be constructed.

We can observe from equation (4) that if we do not demand simultaneous space-time reflections the substitution of the charge e by –e in equation (5) does not reproduce equation (3)

An argument due to Schwinger [10] and enhanced in this article stresses the significance that it is played by the simultaneous space-time reflections to induce charge conjugation.

According to Schwinger the charge of a particle after strong reflections in equations (3) and (5) is given by

$$Q = \frac{1}{c}\int d\sigma_u \langle j_u \rangle = \frac{1}{c}\int d\sigma \langle j_0 \rangle, \tag{6}$$

Where σ is the space-like surface formed by space-time points that cannot be connected by any signal, even by light signals. The charge Q, of equation (6), behaves as a pseudoscalar under time reflections. Hence this transformation interchanges positive and negative charges, and both signs must occur symmetrically in a covariant theory. As a matter of fact, in some cases the requirement of charge symmetry can be substituted by the more incisive demand of invariance under time reflections. However, *this paper goes beyond than that of Schwinger's requirement by demanding that charge symmetry be a corollary of invariance under space-time reflections*. This statement will be proved when the unitary irreducible representations of the Poincare group with space-time reflections be constructed. Wigner [20] proved that time reflection is an antilinear operator by assuming positive energy states [equation (26.3c) in reference 20], then in a circular reasoning circumvent negative energy states.

Landau also conjectures that simultaneous space-time reflections lead to charge conjugation and the exchange of particles by antiparticles [25]. For a field of particles of spin zero, the wave function is given by

$$\psi(r,t) = \frac{1}{\sqrt{\Omega}}\sum_p u\left\{\hat{a}_p e^{-i(\varepsilon t - pr)} + \hat{b}_p^+ e^{i(\varepsilon t - pr)}\right\} \tag{7}$$

If one replaces t -> -t, and r -> -r ones gets

$$\psi(-r,-t) = \frac{1}{\sqrt{\Omega}}\sum_p u\left\{\hat{a}_p e^{i(\varepsilon t - pr)} + \hat{b}_p^+ e^{-i(\varepsilon t - pr)}\right\} \tag{8}$$

On the second quantization formalism, the transition from (7) to (8) leads to the transformation of creation and annihilation operators. This transformation consists in the permutation of $\hat{a}_p$ by $\hat{b}_p^+$, that is

$$\hat{a}_p \rightarrow \hat{b}_p^+, \quad \hat{b}_p^+ \rightarrow \hat{a}_p, \tag{9}$$

In the transformation above, the space inversion changes the momentum, but time inversion reverses it again. The total effect leaves the momentum unchanged. According to this effect the operators with states of equal momentum, transform between themselves. Therefore, the inversion of time changes the future by the past, that is, transforms the creation of a particle by its annihilation. Furthermore, in equation (8) the particles and antiparticles operators are exchanged. That is, the space-time inversion invariance of equation (7) induces the exchange of particles by antiparticles given by equation (9). In other words, *the simultaneous space-time inversion* induces charge conjugation, and the exchange of particles with negative energy by particles of positive energy. This is also true for particles with spin different than zero, because time inversion induces a change on the spin of the particles. This statement will be proved in section 3.1.

If space-time inversion is applied to the Maxwell's equations the same conclusion is reached regarding the interchange between negative and positive charge. Therefore CT is only one transformation, instead of two. CP must be accompanied of the T transformation; otherwise it does not have any physical meaning. A similar argument is valid for PT.

Under the light of the uncertainty principle the arguments against the negative energy states are untenable. Since a particle must reach a speed greater than that of the light to reverse the time direction and acquire a negative mass. Classically this event is impossible, but in quantum mechanics it is totally feasible. Therefore, the potential barrier that has to overcome a particle to be seen as an antiparticle is closely related with the impossibility to reach the speed of light. Let us picture this phenomenon with Klein's paradox. Assume a potential of the form

Region I: $E - V = c\sqrt{|\vec{p}|^2 + m_0^2 c^2} > m_0 c^2$ (10)

Region II: $m_0 c^2 > E - V = c\sqrt{|\vec{p}|^2 + m_0^2 c^2} > -m_0 c^2$

Region III: $-m_0 c^2 > E - V = c\sqrt{|\vec{p}|^2 + m_0^2 c^2}$,

The Klein paradox's argument is useful to visualize the violation of causality on the neighborhood of the light-cone, and to understand the creation of antiparticles when those particles move space-time distances on the order of their Compton wavelength near to the light-cone boundaries. It can be found a very good discussion in reference [21].

According to Klein's paradox, the particle of mass *m* can tunnel from Region I to Region III, in fact the particle, first, has to tunnel to Region II before decaying to Region III. In Region II the particle possess an imaginary mass, $p_\mu p^\mu = E^2 - |\vec{p}|^2 = M^2 < 0$, and hence an imaginary momentum. During this time, the particle is "off-mass-shell," the particle exists as a virtual particle and it is responsible for force transmissions. When finally the particle reaches Region III, passes from a virtual to a real existence, with its mass and energy negatives [16]. To pass from Region I to Region II or III is not a "downhill" transition, since the particle has to reach the speed of light, when moves from Region I to Region II, that is, it has to undergo an infinite acceleration. But an infinite acceleration is classically impossible. However, Heisenberg's uncertainty principle makes this event feasible.

Klein's paradox disappears when the wave packet of the particle is spread out over a distance greater than its Compton wavelength. Since the negative energy states are significant for distances on the order of the Compton wavelength of the particle. According to Heisenberg's principle, if one tries to observe a particle with a very good resolution, one must perturb it with at least an energy-momentum equal its rest mass. As a result of this perturbation oscillations of time could occur, generating antiparticles. For an illuminating discussion see reference [21], pp 45-47 and pp 60-63.

Let us discuss the gravitational collapse under the light of Heisenberg's uncertainty principle as it is given in eq. (2).
According to references 19 and 29, if a cooling star does not reach the equilibrium as a white dwarf or a neutron star, that is, its mass during the thermonuclear evolution does not drop below the Chandrasekhar or Oppenheimer-Volkoff limits, it will collapse; reaching a state of infinite proper energy density in a finite time. However, the uncertainty principle, given by equation (2), would rule out the gravitational collapse. Since a great activity of creation and annihilation of particles-antiparticles would take place when the space-time separation between two particles is on the order of their Compton wavelength. According to Oppenheimer and Volkoff [29] on pages 380-381, there would be only two possible answers to the question of the "final" behavior of very massive stars: *either the equation of state (derivated from Fermi statistics, eq. 11 in reference [29], for a degenerate Fermi gas) that they used fails or the star collapses to form a black hole*. The first answer is the correct, since the particles that obey the Fermi statistics fails for distances shorter than the Compton wavelength of the particles. Therefore, a gravitational collapse would be quantum mechanically inconsistent because the uncertainty principle predicts the creation of particle-antiparticle pairs before such collapse takes place. The neutron stars beyond of $M_{max} = 0.76\odot$, and radius less than $R_{min} = 9.42$km collapses. However, a simple calculation shows that $M_{max}$ is the double of the mass of a Fermi gas of neutrons that can fit in a sphere of $R_{min}=9.42$km. The radius of a neutron was taken on the order of its wavelength. Therefore, these neutrons are wandering around the light-like cone frontier, moving backwards in time (antineutrons). A burst of high-energy gamma rays and π-mesons are emitted for these massive stars, $M>M_{max}$, due to neutron-antineutron annihilation. A more detailed analysis of this event will yield the thermodynamics of massive stars. There is a possibility that a strong interaction among quarks occurs during the contraction and another possible state of equilibrium for the stars could take place. We should search the universe for burst of gamma, and π-mesons instead of the so-called "black holes". Therefore, *the concept of black hole is purely classical* without any support by quantum mechanics.

The thermodynamics of this unstable stars will be similar that the one already discussed by S. Hawking [27].

In this paper the irreducible representations of the Poincare are reviewed to include simultaneous time-space reflections. The irreducible representations of the Poincare group with simultaneous reflection of time and space will be constructed. Then, negative energy states for elementary particles will be naturally generated, enhancing Landau's extended concept of parity conservation, from CP to CPT symmetry, and enforcing Einstein's space-time interpretation.

It is also conjectured in this paper that all the regions formed by the light cone have a physical meaning and that these regions describe the elementary particles. Lamb's shift is an example of the physical existence of one of those regions [21].

The possible existence of dark matter could be a consequence of the negative mass value of the four-dimensional mirror particles reflection. Massless particles have an infinite Compton wavelength, therefore their particles and antiparticles appear in the same proportion. Photons and antiphotons would appear in the same amount. Neutrinos would appear in nature in asymmetric proportion, since these are not massless. At distances compared to the Compton wavelength of system of particles the value of the entropy of a close system could decrease due to the arrow of time that can be reversed.

## 3. Unitary Irreducible Representations of Poincare Group with Simultaneous Space-Time Reflections

The unitary irreducible representations of the extended group of Poincare have been constructed by constraining from the onset the energy values to be positive as a consequence of this assumption time reversal operator is antilinear. Later the antilinear property of the time reversal operator is utilized to prevent the existence of negative energy states for elementary particles [22][23]. In this paper, however the opposite approximation is proposed. The quantum negative energy states are accepted on the basis that the creation of an antiparticle (particle of negative energy) is briefly allowed by an interval of time ruled by the uncertainty principle; while the particle moves around the light cone boundaries that separate time-like from space-like regions. On the neighborhood of this boundary the probability of tunneling is greater than zero. A collapse of particles toward negative energy states is classically prohibited by the impossibility that a particle can exceed the speed of light. However, in quantum mechanics is briefly permitted by the Heinsenberg uncertainty principle (see eq. (2)); as it is discussed above. If this phenomenon occurred continuously, the propagation speed of the information between two objects would be infinite. Furthermore, tunneling has a low probability of occurrence, and the mean lifetime of an antiparticle is in general short, since this is immediately absorbed by its particle to transform into gamma rays.

The classical way to construct the unitary irreducible representations of the Poincare group is by using the technique of the little group. Let us consider the orbits or regions where the magnitude of the four vector $ŝ^2 = x^2+y^2+z^2-c^2t^2$ is: zero, with all its components equal to zero; zero with its components different of zero; greater than zero, and less than zero. To obtain these unitary irreducible representations we will use the two dimensions

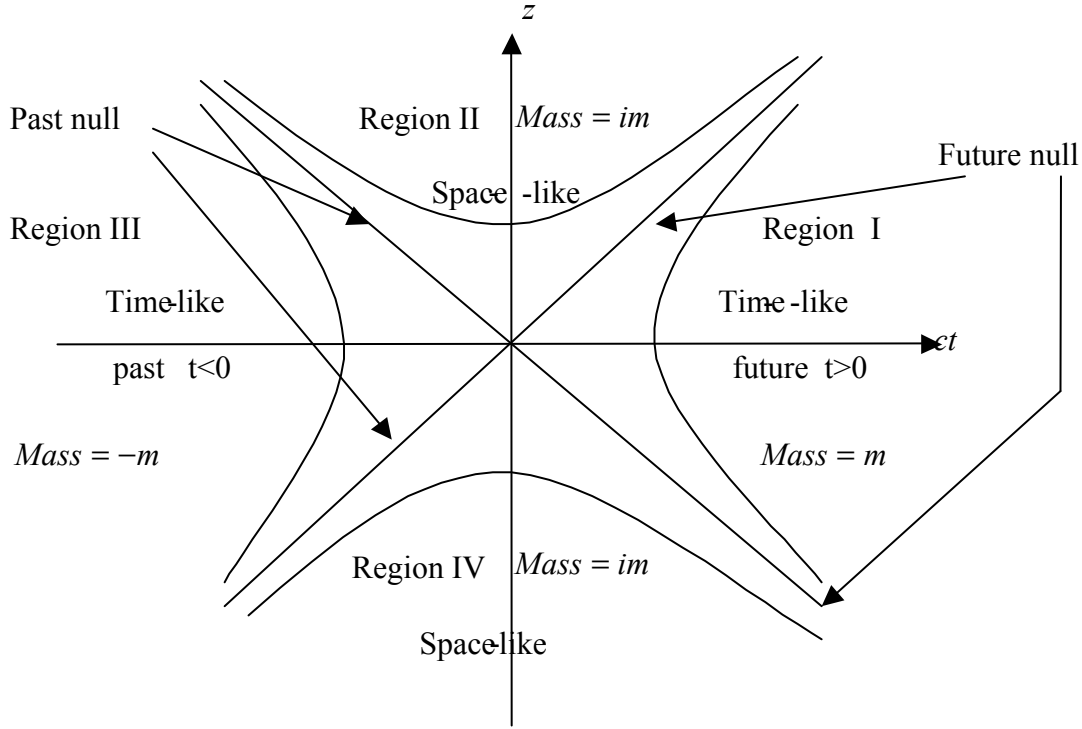

Figure 4

space-time t-z of Figure 4, in the momentum representation. Therefore the six regions to be discussed are

1. $\hat{s}\cdot\hat{s} < 0$, space-like

2. $\hat{s}\cdot\hat{s} > 0$, $t > 0$, future time-like

3. $\hat{s}\cdot\hat{s} > 0$, $t < 0$, past time-like

4. $\hat{s}\cdot\hat{s} = 0$, $\hat{s} \neq 0$, $t > 0$, future null

5. $\hat{s}\cdot\hat{s} = 0$, $\hat{s} \neq 0$, $t < 0$, past null

6. $\hat{s} = 0$, the zero vector

The representation for the Poincare group with simultaneous space-time inversions will be constructed following any of the references [22][23][11].

Let us choose the vectors $|\hat{k}\rangle$ as the basis vectors of an irreducible representation $T^{(k)}$, of the subgroup of translations of the Poincare group. Then the basis vectors $|\hat{k}'\rangle$, with $\hat{k}' = W\hat{k}$, and $W$ a Lorentz transformation, lie in the same representation. That is, $k'$ and $k$ are in the same region of the space-time; these vectors have the same "length", (k',k') = (k,k).

$$T(\hat{u})T(W)|\hat{k}\rangle = T(W)|\hat{k}\rangle \exp(W\hat{k},\hat{u}) \tag{11}$$

The vectors $T(W)|\hat{k}\rangle$ transform according to the irreducible representation $T^{(Wk)}$, of the subgroup of translations.

Let us start our analysis with the future time-like cone or region 2. By choosing the time-like four-vector $\hat{k}_0 = (0,0,0,k)$ with $k>0$, it is found that the little group, corresponding to the orbit of the point $(0,0,0,k)$, is the group of rotations in three dimensions SU(2). Therefore, the representation is labeled by $\hat{k}_0$ and by an irreducible representation label of the three dimensional rotational group SU(2). Starting from the 2s + 1 basis vectors of the little group SU(2) the bases vectors $|\hat{k}sm_s\rangle$, of an irreducible representation of the Poincare group, are generated by a pure Lorentz transformation (a boost) which carries $\hat{k}_0$ into $\hat{k}$, that is $W_k \hat{k}_0 = \hat{k}$. The basis vectors of the representation are given by

$$|\hat{k}sm_s\rangle = T(W_k)|\hat{k}_0 sm_s\rangle. \tag{12}$$

This group operation preserves the "length" of the vector $\hat{k}_0$, that is $(k,k)=(k_0,k_0)$. By applying a Lorentz transformation followed by a translation to equation (12) can be proved that the vectors $|\hat{k}sm_s\rangle$ furnish a unitary irreducible representation of the Poincare group

$$T(\hat{u})T(W)|\hat{k}sm_s\rangle = |\hat{k}'sm_s'\rangle D_{m_s'm_s}(W_{k_0}) \exp(\hat{k}',\hat{u}). \tag{13}$$

Where $\hat{k}' = W\hat{k}$, and $(W\hat{k},W\hat{k}) = (\hat{k},\hat{k})$. $D_{m_s'm_s}(W_{k_0})$ are the unitary irreducible representations of the little group SU(2), $W$ is an arbitrary Lorentz transformation. The representations are unitary because the generators of the group are unitary, and irreducible because the bases vectors of the representations are generated from a single vector $|\hat{k}_0 sm_s\rangle$, with linear momentum equal to zero. Their equivalent unitary irreducible representations are denoted by $P^{(k_0,s)}$. These representations have the same length of the four-vector $\hat{k}$ than that of $\hat{k}_0$. The set of inequivalent representations with different magnitude of the four-vector $\hat{k}$ are denoted by $P^{(k,s)}$.

Let us construct the unitary irreducible representations of the Poincare group with simultaneous space-time reflections. First note that if the vectors $|\hat{k}\rangle$ are the bases of an irreducible representation of the translation group, and $I$ is the space-time reflection operator, then

$$T(\hat{u})T(I)|\hat{k}\rangle = T(I)T(I\hat{u})|\hat{k}\rangle = T(I)|\hat{k}\rangle \exp(I\hat{k},\hat{u}) \tag{14}$$

Therefore, the vectors $T(I)|\hat{k}\rangle$ transform according to the $T^{(Ik)}$ representation of the translation subgroup of the Poincare group. But, the operator $I$ reverses, the space and time components of $\hat{k}$ that is, the energy $E = \hbar c k_t$ becomes negative. On the rest frame of the $P^{(k,s)}$ representation, the rest mass, $m_0 = \dfrac{\hbar I k_0}{c} = -\dfrac{\hbar k_0}{c}$, of such a particle would also be negative. There is nothing to prevent this from happening, since $m_0^2$ is a Poincare invariant [28], but not $m_0$. That is, if one observer sees a particle with rest mass positive another observer on an inertial frame could see the same particle with negative rest mass. An electron with negative rest mass and energy moving backwards in space and time could be interpreted as its antiparticle moving forward in space and time with positive rest mass and energy. This phenomenon could by explained by constructing the unitary irreducible representations of the Poincare group with simultaneous space-time inversions.

To construct the unitary irreducible representations of the full Poincare group one applies the space inversion followed by time inversion to the basis vectors $|\hat{k}sm_s\rangle$ of $P^{(k,s)}$. These basis vectors were generated from the special basis vector $|\hat{k}_0 sm_s\rangle$, where $\hat{k}_0 = (0,0,0,k)$. The space inversion $I_s$ leaves $\hat{k}_0$ invariant, and commutes with the generators of the little group SU(2). Therefore, the irreducible representations of the group, $Z_2 \times SU(2)$, can be labeled by the labels of the rotations, and reflection groups, namely the spin s, and the parity η. If one applies space inversions to the basis vector $|\hat{k}_0 sm_s \eta\rangle$ with definite parity label η = ±, one obtains

$$T(I_s)|\hat{k}_0 sm_s \eta\rangle = |\hat{k}_0 sm_s \eta\rangle \eta \tag{15}$$

On the rest frame the parity, η, is an eigenvalue of the basis vector $|\hat{k}_0 sm_s \eta\rangle$.

Now, Let us start to construct the inequivalent unitary irreducible representations of the Poincare group with space inversions $P^{(k,s,\eta)}$. By applying these inversions to the basis vectors $|\hat{k}sm_s \eta\rangle$, given by equation (12), and including the parity label η one obtains

$$T(I_s)|\hat{k}sm_s\eta\rangle = T(I_s)T(W_k)|\hat{k}_0 sm_s\eta\rangle = T(W_{-k})T(I_s)|\hat{k}_0 sm_s\eta\rangle =$$
$$T(W_{-k})|\hat{k}_0 sm_s\eta\rangle\eta = |I_s\hat{k}sm_s\eta\rangle\eta \tag{16}$$

The Lorentz boost $W_{-k}$ changes the direction of space components of $\hat{k}$. Equation (16) shows that the basis vector $|\hat{k}sm_s\eta\rangle$ is an invariant space under space reflections; the action of the space inversion operator on a general basis vector leads to another basis vector. Hence, these bases vectors yield a representation for the Poincare group with space inversions.

If one considers the simultaneous action of the time-space inversions on a general basis vector one obtains

$$T(I)|\hat{k}sm_s\eta\rangle = T(I)T(W_k)|\hat{k}_0 sm_s\eta\rangle = T(W_k)T(I_t)T(I_s)|\hat{k}_0 sm_s\eta\rangle =$$
$$T(I_t)T(W_{-k})|\hat{k}_0 sm_s\eta\rangle\eta = T(I_t)|I_s\hat{k}'sm_s\eta\rangle\eta = |I\hat{k}'sm_s\eta\rangle\eta \tag{17}$$

The action of the full inversion operator on a general basis vector $|\hat{k}sm_s\eta\rangle$, which belongs to the unitary irreducible representation $P^{(k,s,\eta)+}$ generated from the vector $\hat{k}_0 = (0,0,0,k)$, leads to a basis vector in the $P^{(k,s,\eta)-}$ representation generated from the vector $I\hat{k}_0 = (0,0,0,-k)$. But, these two representations of the full Poincare group are equivalents. Therefore, the action of time inversion generates negative energy states, and also induces charge conjugation, as we will show it below.

**3.1 Time Inversion and Charge Conjugation**

If one applies linear time reflections to the basis vector $|\hat{k}_0 sm_s\eta\rangle$, one finds that

$$T(I_t)|\hat{k}_0 sm_s\eta\rangle = |I_t\hat{k}_0 sm_s\eta\rangle \tag{18}$$

Since, from equation (14) $T(I_t)|\hat{k}_0 sm_s\eta\rangle$ transforms according to the $T^{(I_t k_0 s)}$ representation of the subgroup of translations. That is, the basis vector $|\hat{k}_0 sm_s\eta\rangle$ with $\hat{k}_0 = (0,0,0,k)$ from the $P^{(k_0,s,\eta)+}$ representation is taken into the basis vector $|I_t\hat{k}_0 sm_s\eta\rangle$ with $I_t\hat{k}_0 = (0,0,0,-k)$ of the $P^{(k_0,s,\eta)-}$ representation. We will show that in the $P^{(k_0,s,\eta)-}$ representation, time reflections prevent that SU(2) commutes with $Z_2$. In spite of SU(2) does not commute with $Z_2$, in the $P^{(k_0,s,\eta)-}$ representation, still η is a label for the full

Poincare group representation. Since the Casimir operators of the full Poincare groups are $P^2$, $W^2$, and I.

If the vector $|\hat{k}\rangle$ is any vector that transforms according to a representation of the group of translations, then

$$\hat{P}|\hat{k}\rangle = -i\hat{k}|\hat{k}\rangle \qquad (19)$$

Hence, in the $P^{(k_0,s,\eta)+}$ representation

$$\hat{P}|\hat{k}_0 sm_s\eta\rangle = -ik|\hat{k}_0 sm_s\eta\rangle \qquad (20)$$

While in the $P^{(k_0,s,\eta)-}$ representation

$$\hat{P}|I_t\hat{k}_0 sm_s\eta\rangle = +ik|I_t\hat{k}_0 sm_s\eta\rangle \qquad (21)$$

Then, time reflection induces a change on the energy sign, that is

$$I_t P_0 I_t^{-1} = -P_0 \qquad (22)$$

That is, time inversions do not commute with $P_0$.

The Pauli-Lubanki four vector components on the rest frame in the $P^{(k_0,s,\eta)+}$ representation is

$$W_q|\hat{k}_0 sm_s\eta\rangle = -kJ_q|\hat{k}_0 sm_s\eta\rangle, \text{ and}$$

$$W_t = 0, q = x, y, z \qquad (23)$$

Now, in the $P^{(k_0,s,\eta)-}$ representation, on the orbit of the vector $I_t\hat{k}_0 = (0,0,0,-k)$ yields

$$W_q|I_t\hat{k}_0 sm_s\eta\rangle = +kJ_q|I_t\hat{k}_0 sm_s\eta\rangle, \text{ and}$$

$$W_t = 0, q = x, y, z \qquad (24)$$

So that, time inversions do not commute with the Pauli-Lubansky four-vector in the larger space composed by $P^{(k_0,s,\eta)+}$ and $P^{(k_0,s,\eta)-}$.

$$I_t W_q I_t^{-1} = -W_q, \qquad (25)$$

Therefore, from equation (24) one obtains

$$I_t J_q I_t^{-1} = -J_q \qquad (26)$$

That is, time reversal induces an inversion on the direction of a rotation and changes the sign of the rest energy (rest mass) of the particle. Then, by using equation (23) one gets

$$J_z I_t |\hat{k}_0 s m_s \eta\rangle = -I_t J_z |\hat{k}_0 s m_s \eta\rangle = I_t |\hat{k}_0 s m_s \eta\rangle (-m_s) \qquad (27)$$

The vector $I_t |\hat{k}_0 s m_s \eta\rangle$ transforms like $-m_s$, under rotations about the z-axis. Therefore from equation (26) for a general rotation, we get

$$T(R(\theta))T(I_t)|\hat{k}_0 s m_s \eta\rangle = T(I_t)T(R(-\theta))|\hat{k}_0 s m_s \eta\rangle =$$

$$T(I_t)\sum_{m_s'}|\hat{k}_0 s m_s' \eta\rangle D^{(s)-1}_{m_s' m_s}(\theta) = T(I_t)\sum_{m_s'}|\hat{k}_0 s m_s' \eta\rangle \widetilde{D^{(s)*}_{m_s' m_s}}(\theta) \qquad (28)$$

Hence, the vectors $T(I_t)|\hat{k}_0 s m_s \eta\rangle$ transform like the transpose conjugate complex, $D^{(s)*}_{m_s' m_s}$, of the representation $D^{(s)}_{m_s' m_s}$ of the unitary irreducible representation of SU(2). This *fact explains why it is necessary to conjugate and transpose the wave equation of a particle to describe its antiparticle*. Thus, time reflection induces negative energy states and these states induce charge conjugation. If a mirror reflection is a symmetry transformation, then this reflection must be accompanied by simultaneous space-time inversions, since the intrinsic parity label is generated by space reflection. Equations (6) and (8) clearly show this fact.

Due to the fact that one has to conjugate and transpose, time reversal acquires the properties of an antilinear operator. And since the character of the representations of the three-dimensional rotation group is real, the representations, $D^{(s)*}_{m_s' m_s}$ and $D^{(s)}_{m_s' m_s}$, are equivalent and those representations can be reached one from the other by a change of basis. We apply this transform to the transpose of the transpose of complex conjugated representation. On the new basis the basis vector is given by

$$\left|\hat{k}_0 s m_s \eta\right\rangle_{\substack{new \\ basis}} = \sum_{m_s'} \delta^{m_s'}{}_{-m_s}(-)^{s-m_s'} \left|\hat{k}_0 s m_s' \eta\right\rangle \tag{29}$$

Since the matrix $D^{(s)}$ transforms into the matrix $D^{(s)*}$ according to a set of similarity transformations. As a matter of fact we are mapping a vector over its dual. Then, one has to define

$$\widetilde{D^{(s)*}_{m_s' m_s}} = F^{-1} D^{(s)}_{m_s' m_s} F^1 \tag{30}$$

where the transformation matrix $F$ is given by

$$F = \delta^{m_s'}{}_{-m_s}(-)^{s-m_s} \tag{31}$$

By applying the linear time reversal operator, on the new basis vector, to the general basis vector $\left|k s m_s \eta\right\rangle$ which, is defined by equation (12), one gets

$$T(I_t)\left|\hat{k} s m_s \eta\right\rangle = \sum_{m_s'} T(I_t)\left|\hat{k} s m_s' \eta\right\rangle \delta^{m_s'}{}_{-m_s}(-)^{s-m_s'} = \left|I_t \hat{k} s - m_s \eta\right\rangle (-)^{s+m_s} \tag{32}$$

Under the action of the time reversal operator a particle on the rest frame reverses its energy, and spin.

We define the scalar product according to reference [11] by

$$\left|\psi\right\rangle = \sum_{m_s} \int \psi_{m_s}(\hat{k}) \left|\hat{k} s m_s\right\rangle k_t^{-1} dk \tag{33}$$

Therefore applying the bra-vector of (32) to equation (33) one gets

$$\psi'_{m_s}(\hat{k}) = (-1)^{s-m_s} \psi^*_{-m_s}(I_t \hat{k}) \tag{34}$$

Here, the conjugation is a consequence of the bra-vector. Applying the space reversal operator to the general basis vector $\left|k s m_s \eta\right\rangle$, one obtains

$$T(I_s)\left|\hat{k} s m_s \eta\right\rangle = T(W_{-k})T(I_s)\left|\hat{k}_0 s m \eta_s\right\rangle = T(W_{-k})\left|\hat{k}_0 s m_s \eta\right\rangle \eta = \left|I_s \hat{k} s m \eta_s\right\rangle \eta \tag{35}$$

The momentum is reversed and the particle acquires a definite parity. In order to generate the basis vectors of the representation for the Poincare group with simultaneous space-time reflections, we apply the time reversal operator to equation (16)

$$T(I)\left|\hat{k}sm_s\eta\right\rangle = T(I_t)T(I_s)T(W_k)\left|\hat{k}_0 sm_s\eta\right\rangle = T(I_t)T(W_{-k})\left|\hat{k}_0 sm_s\eta\right\rangle\eta$$
$$= T(I_t)\left|I_s\hat{k}sm_s\eta\right\rangle\eta = \left|I\hat{k}s - m_s\eta\right\rangle\eta(-)^{s+m_s} \tag{36}$$

Therefore, in this formulation the simultaneous action of space-time inversions on a general basis vector (particle) of the representation reverses the energy, momentum, and spin (antiparticle). The space inversion furnishes a definite parity, η, to the elementary particle described by the unitary irreducible representation. The elementary particle violates the causality principle because it is represented in the negative energy sector of the light cone. If it has enough energy it can be observed as a particle moving backwards in space and time (antiparticle). Therefore, the C, T, P, CT, CP, PT cannot conserve separately, but CPT. In my opinion not all the symmetries that would enhance the full Poincare group are known. Then, it is quite possible that exists a violation of CPT for physical phenomena that requires more large symmetries. The full group of diffeomorphisms (space-time reflections included) is a larger symmetry than that of the full Poincare group; therefore CPT could be violated by quantum gravity. By the same token supersymmetry, supergravity, and superstrings could violate CPT.

From equation (36), we get

$$T^2(I)\left|\hat{k}sm_s\eta\right\rangle = \left|\hat{k}sm_s\eta\right\rangle(-)^{2s} \tag{37}$$

Hence, $T^2 = 1$ for integer spin, and $T^2 = -1$ for particles of half integer spin.

### 4. Particles of Zero Mass

For massless particles, or the future null and past null regions 4 and 5, the unitary irreducible representations of the Poincare group are labeled by unitary irreducible representations of its little group. The little group of that of Poincare group in these regions is the Euclidean group $E^{(2)}$ in two-dimensions, one will denote its elements by $W_{k_0}$. The $W_{k_0}$'s are the transformations that leave invariant the four-vector $\hat{k}_0$. The orbit of the vector $\hat{k}_0 = (0,0,1,1)$ is the light cone surface. Let $W_k$ be a Lorentz transformation that takes $\hat{k}_0$ into $\hat{k}$, that is $W_k\hat{k}_0 = \hat{k}$; then a general vector can be written by

$$\left|\hat{k}m\right\rangle = T(W_k)\left|\hat{k}_0 m\right\rangle. \tag{38}$$

It can be proved that $W_{k'}^{-1}WW_k$ is in the little group, that is $W_{k_0} = W_{k'}^{-1}WW_k$, then applying a general Lorentz transformation $W$ followed by a translation $\hat{u}$ one obtains.

$$T(\hat{u})T(W)|\hat{k}m\rangle = |\hat{k}'m\rangle \exp(ik',\hat{u})\exp(-im\theta) \quad (39)$$

The $E^{(2)}$ representations with continuous spin were ignored in this revision. We took on consideration only the representations labeled by the two-dimensional rotation group, that is $m = 0, \pm\frac{1}{2}, \pm 1..., etc.$

By enlarging the Poincare group to include space-time inversions the little group of the vector $\hat{k}_0 = (0,0,1,1)$ $\widetilde{E}^{(2)} = Z_2 \times E^{(2)}$. Hence, the "full" Poincare group is labeled by η = ±1 and a label of $SO(2)$, because the continuous spin representations of $E^{(2)}$ will not be considered in these regions of the light cone, then $\widetilde{E}^{(2)}$ is labeled by $SO(2)$. If one applies space-time inversion over the basis vector

$$|\hat{k}m\rangle = T(W_k)|\hat{k}_0 m\rangle \quad (40)$$

Where $W_k \hat{k}_0 = \hat{k}$, furthermore if one notices that under space inversions

$$(W_q - imp_q)T(I_s)|\hat{k}m\rangle = T(I_s)(W_q + imp_q)|\hat{k}m\rangle = 0 \quad (41)$$

$$(W_t - imp_t)T(I_s)|\hat{k}m\rangle = T(I_s)(W_t + imp_t)|\hat{k}m\rangle = 0 \quad (42)$$

Hence,

$$T(I_s)|\hat{k}m\rangle = |I_s\hat{k} - m\rangle \quad (43)$$

and

$$T(I_s)|\hat{k} - m\rangle = |I_s\hat{k}m\rangle \quad (44)$$

Therefore, these vectors belong to the $P^{(0,m)} + P^{(0,-m)}$ representation, contrary to the representation $P^{(k,m)}$ these representations do not carry spin, but helicity. Therefore, time reversal does not induces a change in sign of the angular momentum *J,* as it does in equation (24).

Also, if one notices that

$$(W_q + imp_q)T(I_t)T(I_s)|\hat{k}m\rangle = T(I_t)(W_q - imp_q)T(I_s)|\hat{k}m\rangle = 0$$

$$(W_t + imp_t)T(I_t)T(I_s)|\hat{k}m\rangle = T(I_t)(W_t - imp_t)T(I_s)|\hat{k}m\rangle = 0 \qquad (45)$$

Hence,

$$T(I_t)T(I_s)|\hat{k}m\rangle = |I_t I_s \hat{k}m\rangle \qquad (46)$$

Therefore, under space-time inversions the vector $T(I)|\hat{k}m\rangle$ transforms according to the representation $T^{(Ik)}$. Additionally it can be proved that such a vector transforms in the same manner under translations. Now, according to the equations

$$(W_q + imp_q)|\hat{k}m\rangle = 0$$

$$(W_t + imp_t)|\hat{k}m\rangle = 0. \qquad (47)$$

The $|I\hat{k}m\rangle$ vector transforms according to the $P^{(0,m)}$ representation but, with its energy and momentum reversed.

On the $|\hat{k}_0 0\eta\rangle$ representation of the little group of $\hat{k}_0$ one gets

$$T(I)|\hat{k}0\eta\rangle = T(W_k)T(I_t)T(I_s)|\hat{k}_0 0\eta\rangle = |I\hat{k}_0 0\eta\rangle\eta \qquad (48)$$

This equation represents a particle of zero rest mass, and zero helicity, but with a definite parity that changes its momentum and energy under space-time reversal.

We realize that transitions between time-like and space-like states for a "massless" particle are highly probable, since if the four momentum, $p^2 = 0$, is sharply defined, then its four vector position

$$(x_1 - x_2)^2 - (x_1^0 - x_2^0)^2 \sim \frac{\hbar^2}{p^2 \to 0} \qquad (49)$$

is completely uncertain, since its Compton wavelength is undetermined [27]. The massless particle could exist in any place of the light cone. The principle of causality is strongly violated. If one observes a given amount of photons with $p^2 = 0$ is because a big amount of other photons populate the space-like, and the time-like regions. The action, of these dark photons, on the universe should be measurable.

**5. Imaginary Mass Particles**

In the space-like, region 1, the eigenvalue of the square of the four-momentum is the negative of the square of the rest mass of the particle. Therefore, the mass would be imaginary number. No physical interpretation of elementary particles could be associated to this region; however, the square of the four-momentum could have a physical meaning. It turn out that the square of the four-momentum is not equal to the square of the mass for internal lines in a Feynman diagram. It is conjectured in this article that a possible interpretation could be given to the unitary irreducible representations of the "full" Poincare group. Virtual particles would be good candidates to achieve this goal.

The little group of the four-vector $\hat{k}_0 = (0,0,k,0)$ is the SL(2,r) which contains spin representations. These representations where classified independently by Bargmann, Naimark and Gelfand, and Harish-Chandra[23].

It can be shown that a general basis vector is given by

$$\left| \hat{k}m \right\rangle = T(W_k)\left| \hat{k}_0 m \right\rangle = T(R_z(\theta))T(W_{k_1})T(W_{k_3})\left| \hat{k}_0 m \right\rangle \tag{50}$$

If we apply the space reversal operator to the four basis vector and consider as the little group the group of reflections and SL(2,R), $Z_2 X SL(2,R)$, one obtains

$$T(I_s)\left| \hat{k}sm_s\eta \right\rangle = \sum_{m'_s} T(I_s)\left| \hat{k}sm'_s\eta_s \right\rangle \delta^{m'_s}_{-m_s}(-)^{s-m'_s} = \left| I_s\hat{k}s - m_s\eta \right\rangle (-)^{s+m_s}, \tag{51}$$

since the relation

$$I_s J_z I_s^{-1} = -J_z, \tag{52}$$

is obtained when one chooses the point $\hat{k}_0 = (0,0,-z,0)$ to construct the space-like representations, by following a similar procedure than that of section 3.1. Applying time reversal to the general basis vector, one finds that the four-vector $\hat{k}$ reverses its spatial components.

$$T(I_t)\left|\hat{k}m\eta\right\rangle = T(W_{-k})T(I_t)\left|\hat{k}_0 m\eta\right\rangle = \left|I_s\hat{k}m\eta\right\rangle\eta \qquad (53)$$

Hence, the action of space-time inversion yields

$$T(I)\left|\hat{k}sm_s\eta\right\rangle = T(I_t)T(I_s)T(W_k)\left|\hat{k}_0 sm_s\eta\right\rangle = T(I_s)T(W_{-k})\left|\hat{k}_0 sm_s\eta\right\rangle\eta$$
$$= T(I_s)\left|I_s\hat{k}sm_s\eta\right\rangle\eta = \left|\hat{k}s - m_s\eta\right\rangle\eta(-)^{s+m_s} \qquad (54)$$

The vectors of the discrete series $D^+$ transform into the $D^-$ discrete series, and vice versa.

### 6. Vacuum Representations

In region six the little group of the four-vector $\hat{k}_0 = (0,0,0,0)$ is the Lorentz group. Because the Lorentz group is non-compact, its unitary irreducible representations are infinite dimensional. Therefore, the unitary irreducible representations of the Poincare group in this region are the same than that of the Lorentz group. They are labeled by the eigenvalues of the Casimir of the algebra SU(2)XSU(2), that is, the angular momentum on three dimensions u, v. The Lie algebra of the Lorentz group can be reduced to that of the group SU(2)XSU(2). Although these two groups are distinct topologically, the Casimir operators of SU(2)XSU(2) enable one to deduce and label the representation of the Lorentz group. There are two classes of unitary irreducible representations of the Lorentz group, the principal series ($v$ = -iw, w real, $j_0$ = 0,1/2,1,...) and the complementary series (-1 ≤ $v$ ≤ 1; $j_0$ = 0). Because combined space-time transformations commute with the generators of the Lorentz group the states of this representation have a definite parity. Furthermore, vacuum oscillations make these states to undergo transitions to positive and negative energy states as well as imaginary mass (virtual states).

**Conclusion**

The main purpose of this paper was to recover parity conservation symmetry. A general review of the full Poincare group was given. In the process it was discussed a possible entropy decreasing, as well as possible violations of CPT. One also conjectured the existence of negative mass, and the incompatibility of quantum mechanics with a gravitational collapse.
One has proved that the parity is conserved if one reverses simultaneously the space-time coordinates. It has been shown in this paper that what it is called PT is miscalculated because time is forced to be an anti-linear operator by brute force in order to maintain the energy of the elementary particles positive. In our new scheme PT is actually CPT because time inversions induce charge conjugation. Therefore, CP conservation, in the old scheme, does not make any sense, similarly PT. Hence, Yang and Landau were proposing an incomplete symmetry [5] when they suggested reflecting the space coordinates without inverting time, and exchanged particles by antiparticles. The inversion of time induces charge conjugation.

Although many scientists were uncomfortable with the old concept of mirror reflection [7][24], none rational theory were formulated up to the publication of this anuscript. Superstring Theory and other theories based on larger symmetry groups than that of the "full" Poincare group could violate CPT conservation, since the extended symmetries could include an (or some) unknown X, symmetry (symmetries). So, the full symmetry for the elementary particles in this case would be CPTX. A quantum theory of gravity based on the full Einstein group (group of general coordinate transformations plus space-time inversions) could violate CPT.

**Acknowledgments**

I would like to thank George Sudarshan for discussing with me the incompatibility of Bose-Einstein and Fermi-Dirac statistics to explain the creation of antiparticles under the same theoretical frame. I was illuminated by a short discussion with Carlos Castro, both conversations with George and Carlos took place when I was a graduate student in 1992. I am in debt with Yuval Ne'eman to help me to understand the importance of group theory in physics. I would also like to thank Steven Weinberg for giving me the opportunity to be a visiting scholar in his group during the summer of 2002. This paper was inspired by Weinberg's book in Gravitation. However, the author of this article assumes all the responsibility concerning any misleading interpretation. The last acknowledgment, but not the least goes to Roger Pecina for his worthy commentaries.